\documentclass[fleqn,usenatbib]{mnras}

\usepackage{newtxtext,newtxmath}

\usepackage[T1]{fontenc}

\usepackage{graphicx}	
\usepackage{amsmath}	

\defcitealias{Lohmer2001}{L2001}
\defcitealias{Lewandowski2013}{Lew2013}
\defcitealias{Lewandowski2015a}{Lew2015b}

\title[Scattering analysis of single-component pulsars]{The Thousand-Pulsar-Array programme on MeerKAT -- V. Scattering analysis of single-component pulsars}

\author[L. S. Oswald et al.]{
L. S. Oswald$^{1}$\thanks{E-mail: lucy.oswald@physics.ox.ac.uk (LSO)},
A. Karastergiou$^{1,2}$,
B. Posselt$^{1,3}$,
S. Johnston$^{4}$,
M. Bailes$^{5,6}$,\newauthor
S. Buchner$^{7}$,
M. Geyer$^{7}$,
M.~J.~Keith$^{8}$,
M.~Kramer$^{9}$,
A.~Parthasarathy$^{9}$,
D.~J.~Reardon$^{5,6}$,\newauthor
M.~Serylak$^{7}$,
R.~M.~Shannon$^{5,6}$,
R.~Spiewak$^{5,8}$, 
W.~van Straten$^{10}$,\newauthor
V.~Venkatraman~Krishnan$^{9}$
\\
$^{1}$Department of Astrophysics, University of Oxford, Denys Wilkinson Building, Keble Road, Oxford OX1 3RH, UK\\
$^{2}$Department of Physics and Electronics, Rhodes University, PO Box 94, Grahamstown 6140, South Africa\\
$^{3}$ Department of Astronomy \& Astrophysics, Pennsylvania State University, 525 Davey Lab, 16802 University Park, PA, USA\\
$^{4}$CSIRO Astronomy and Space Science, Australia Telescope National Facility, PO~Box~76, Epping NSW~1710, Australia\\
$^{5}$Centre for Astrophysics and Supercomputing, Swinburne University of Technology, Hawthorn, VIC, 3122 Australia\\
$^{6}$ ARC Centre of Excellence for Gravitational Wave Discovery (OzGrav)\\
$^{7}$SARAO, 2 Fir Street, Black River Park, Observatory, 7925, South Africa\\
$^{8}$Jodrell Bank Centre for Astrophysics, Department of Physics and Astronomy, University of Manchester, Manchester M13 9PL, UK\\
$^{9}$Max-Planck-Institut f\"{u}r Radioastronomie, Auf dem H\"{u}gel 69, D-53121 Bonn, Germany\\
$^{10}$Institute for Radio Astronomy \& Space Research, Auckland University of Technology, Private Bag 92006, Auckland 1142, New Zealand\\
}

\date{Accepted XXX. Received YYY; in original form ZZZ}

\pubyear{2020}

\begin{document}
\label{firstpage}
\pagerange{\pageref{firstpage}--\pageref{lastpage}}
\maketitle

\begin{abstract}
We have measured the scattering timescale, $\tau$, and the scattering spectral index, $\alpha$, for 84 single-component pulsars. Observations were carried out with the MeerKAT telescope as part of the Thousand-Pulsar-Array programme in the MeerTime project at frequencies between 0.895 and 1.670~GHz. 
Our results give a distribution of values for $\alpha$ (defined in terms of $\tau$ and frequency $\nu$ as $\tau\propto\nu^{-\alpha}$) for which, upon fitting a Gaussian, we obtain a mean and standard deviation of $\langle\alpha\rangle = 4.0 \pm 0.6$. This is due to our identification of possible causes of inaccurate measurement of $\tau$, which, if not filtered out of modelling results, tend to lead to underestimation of $\alpha$. The pulsars in our sample have large dispersion measures and are therefore likely to be distant. We find that a model using an isotropic scatter broadening function is consistent with the data, likely due to the averaging effect of multiple scattering screens along the line of sight. Our sample of scattering parameters provides a strong data set upon which we can build to test more complex and time-dependent scattering phenomena, such as extreme scattering events.
\end{abstract}




\section{Introduction}

A commonly observed signature of the effect of the interstellar medium (ISM) upon the radio emission from pulsars is the scattering of the radio flux. Regions of cold ($<10,000$K), dense plasma along the line of sight, which can be approximated as one or multiple thin screens, cause some of the radio flux arriving at the observer to be delayed with respect to the direct line of sight. Observationally, this results in a scattered pulse profile with a characteristic exponential scattering tail \citep{Cronyn1970}, which can be explained assuming isotropic scattering screens, for which the scattering angles have no preferred direction. 
Measurements of the scattering properties of pulsars reveal information about the structure of the ISM and allow us to disentangle these effects from the intrinsic properties of the pulse shape.

In the context of isotropic scattering with an exponential transfer function the key parameter is the scattering timescale, $\tau$.
It is observed that $\tau$ evolves with frequency $\nu$ according to the power law relationship
\begin{equation}
    \tau \propto \nu^{-\alpha},
    \label{eq:taufreq}
\end{equation}
where $\alpha$ is the scattering spectral index. The simplest thin screen scattering model predicts $\alpha$ = 4 (e.g. \citealp{Cronyn1970, Lang1971a}), whilst a medium exhibiting Kolmogorov turbulence would have a value of 4.4 (e.g. \citealp{Lee1976a, Rickett1977a}). The simplified spectrum of turbulence is described by a power law for wavenumbers $q$, for values of $q^{-1}$ that lie well within inner and outer fluctuation scales ($k_{i}^{-1}$ and $k_{o}^{-1}$ respectively). This is written as $P_{n_{e}}(q) = C_{n_{e}}^{2}q^{-\beta}$ \citep{Rickett1977a}, where $C_{n_{e}}^{2}$ is the proportionality constant, dependent on the electron density. The fluctuation spectral index $\beta$ cannot exceed 4 \citep{Romani1986}. It is related to $\alpha$ by $\alpha = 2\beta/(\beta - 2)$, which leads to $\alpha = 4$ being a theoretical upper limit.  Observationally however, lower values of $\alpha$, i.e. flatter spectra, are commonly seen \citep[see for example][]{Krishnakumar2019}. 
A power law break at low frequencies with the scattering timescale approaching frequency independence, accompanied by a loss of flux, may be attributed to a truncated scattering screen \citep{Cordes2001}. 
Where $q^{-1}$ drops below the inner scale of the plasma turbulence, $k_{i}^{-1}$, the power law description of the turbulence spectrum is no longer a valid simplification. \cite{Lewandowski2013, Lewandowski2015} write that, for a single pulsar, this may lead to a flattening of $\alpha$ at lower frequencies. The effect of this inner scale, which corresponds to the shortest scale length in the scattering material, should also become apparent in the scattered pulse shape, particularly at large delays in the profile tail \citep{Rickett2009}. 
Furthermore, the exact shape of the scattered profile will depend upon the degree of anisotropy in the scattering material. Evidence of such anisotropy has been seen in the parabolic arcs observed in the secondary spectra of scattered pulsar observations (e.g. \citealp{Stinebring2001, Walker2004}) and high resolution mapping of the locations of scintils \citep{Brisken2010, Pen2014}. \cite{Geyer2016b} showed that fitting an isotropic model to simulated anisotropically scattered data could lead to inferring smaller values of $\alpha$.

In the literature, values of $\alpha$ are usually calculated by performing a power law fit to $\tau$ against frequency, where $\tau$ is measured through one of two ways. In the cases where the scatter broadening is evident in the pulse profile, we measure $\tau$ in the time domain. This is done either through forward modelling \citep[e.g.][]{Geyer2017a} or through deconvolution analysis, such as the {\sc{clean}} algorithm \citep{HOGBOM1974, Bhat2003}. Scattering surveys that have employed forward modelling, and are thus directly comparable to the work presented here, are as follows: \cite{Cordes1985, Lohmer2001, Lohmer2004, Kuzmin2007, Lewandowski2011, Lewandowski2013, Lewandowski2015, Lewandowski2015a, Krishnakumar2017, Geyer2017a} and \cite{Krishnakumar2019}. The {\sc clean} algorithm has been employed by, for example, \cite{Bhat2004a} and \cite{Kirsten2019}. A complementary technique is to infer $\tau$ from the scintillation bandwidth $\delta\nu$ \citep{Cordes1985}. This is generally done using the equation $2\uppi\tau\delta\nu = C_{1}$, where $C_{1}$ is a factor with a value that depends on the electron density wavenumber spectrum and distribution of scattering material. When transforming observational values it is usually set to $C_{1} = 1$, the solution for a thin screen, and other values are presented in \cite{Cordes1998}. This technique is relevant for observations where $\tau$ is too small to be measured directly, generally at higher frequencies. The inverse is also true: where $\tau$ is large, $\delta\nu$ will be too small to be measurable.

The scattering strength, which determines the choice of methodology used to measure $\tau$ and $\alpha$, is expected to be correlated with distance for a given pulsar, added to which is a further stochastic element due to there generally being a small number of scattering screens along the line of sight. 
Modelling the scattering material as a power-law electron density spectrum with a Kolmogorov distribution of irregularities, the relationship between scattering timescale and DM is expected to be $\tau \propto C_{n_{e}}^{2}\nu^{-4.4}\rm{DM}^{2.2}$ \citep{Romani1986, Cordes1998, Krishnakumar2015}. 

The new MeerKAT telescope \citep{Bailes2020b} provides high sensitivity in the observing band 856--1712~MHz. The high quality pulse profiles and the broad bandwidth are 
suitable for a survey of scattered pulsars where measurements of $\tau$ and $\alpha$ can be made using a single instrumental setup. 
This largely eliminates systematic errors arising from measurements of $\tau$ made using multiple receivers and backends at different frequencies. Our survey of scattered pulsars is part of the work of the Thousand-Pulsar-Array (TPA) \citep{Johnston2020a}. This is an observing project that is being carried out as part of MeerTime, a large scale project to observe known pulsars using the MeerKAT telescope \citep{Bailes2020b}. 

The TPA programme will obtain an overview of the properties of a large sample of the observed pulsar population. 
The structure of this paper is as follows: section \ref{sec:proc} describes the observations and data processing. 
Section \ref{sec:model} describes our new time domain methodology to determine $\tau$, which uses a Markov Chain Monte Carlo (MCMC) fit. 
The results of the scattering analysis, both a table of measured scattering parameters and a description of the distributions of these parameters across the population of observed pulsars, are presented in section \ref{sec:results}. We also describe how our observational constraints affect the sample of pulsars for which we are able to accurately measure scattering parameters, and we investigate how the level of covariance between the parameters affects our modelling outcomes, before discussing our results in the context of previous work in section \ref{sec:disc}. Conclusions are summarized in section \ref{sec:conc}. 

\section{Observations and data}
\label{sec:proc}

We selected 205 pulsars from the TPA programme with simple profiles that resemble a single Gaussian component convolved with an exponential function. 
We chose these by eye out of 1164 pulsars observed up until 1st June 2020. The initial sample was deliberately selected to be as broad as possible within these constraints, so that the sample was not biased by visual selection. This allowed us to investigate the limitations of our modelling choices, but required us to discard at later stages those pulsars whose profiles were not well modelled with single Gaussians modified by scattering tails. 
The observing and data processing was carried out as described in \cite{Johnston2020a} and \cite{Serylak2020}. The pulsars were observed in fold mode over multiple epochs using MeerKAT at L-band. These observations were de-dispersed, cleaned of Radio Frequency Interference (RFI) using {\sc CoastGuard} \citep[][see ascl.net/2003.008]{Lazarus2016} and time-integrated to a single Stokes I profile per observation with a resolution of 1024 bins per pulse period $P$. Observations from the early commissioning phase of the MeerKAT telescope had only the band between 895 and 1670~MHz available out of the ultimate full band of 856--1712~MHz. In order to make all our observations directly comparable with each other, we therefore kept the band between 895 and 1670~MHz, divided into 8 sub-bands, in the final data products. 
As originally shown by \cite{Geyer2016b}\footnote{Equations 5 and 6 in \cite{Geyer2016b} have typos, with $f_{m}$ and $f_{c}$ interchanged. As of 4th February 2021, this has been corrected in the arXiv version of the paper, found at https://arxiv.org/pdf/1607.04994.pdf}, assuming a power-law relationship between scattering timescale and frequency, the appropriate frequency associated with a given measurement of $\tau$ is given in terms of the centre frequency, $f_{c}$, and bandwidth, $\delta f$, of the profile as 
\begin{equation}
    f_{m} = 10^{\left[\log_{10}\left(f_{c} + \delta f/2\right) + \log_{10}\left(f_{c} - \delta f/2\right)\right]/2}.
\end{equation}
Where a pulsar has multiple good quality observations taken at different epochs, we aligned and added these observations to produce a single profile with increased signal-to-noise ratio (S/N). We did this through the following method. Within the TPA-project, pulse templates were obtained. Ephemerides were updated when the data showed apparent deviations from the known ones. Employing these ephemerides, we used {\sc tempo2} \citep{Hobbs2006tempo2,Edwards2006tempo2} and {\sc psrchive} \citep{vanStraten2012} to obtain phase shifts between individual observations of a pulsar, which we used to align them in phase before adding.

\section{Modelling scattered pulsar profiles}
\label{sec:model}

\subsection{Method}
\label{sec:method}

We modelled the scattered pulse profile as a single Gaussian component representing the intrinsic emission, convolved with an exponential scattering function. We assumed that the scattering material took the form of a thin screen with isotropic scattering properties. The temporal broadening function was therefore described in terms of the scattering timescale $\tau$ as
\begin{equation}
    \frac{e^{-t/\tau}}{\tau} U(t)
    \label{eq:iso}
\end{equation}
 where the unit step function $U(t)$ constrains the equation to time $t > 0$ \citep{Cordes2001}. We fit for 5 components: the amplitude ($A$), mean ($\mu$) and standard deviation ($\sigma$) of the Gaussian; the scattering timescale ($\tau$); and any DC offset of the profile baseline. Early modelling also investigated the applicability of an extreme anisotropic model, as given by equation 3 in \cite{Geyer2017a}. However, since this was found to be generally less successful at fitting the profiles than the isotropic model, we did not take this further.

We fit each of the 8 sub-bands per pulsar independently. We used the \textit{train $+$ DC} method as described in \cite{Geyer2016b}, updating the fitting process to sample the log-likelihood function over parameter space using the MCMC algorithm {\sc emcee} \citep{Foreman-Mackey2013}. We imposed flat priors, with the only constraint that the parameters had to be physical, in order to allow the fit to converge freely. This meant we constrained $A$ and $\tau$ to be positive, and $0 < \mu < 1024$ and $0 < \sigma < 1024$, where 1024 is the number of bins across the pulse period. We ran the MCMC for either 20,000 steps per frequency, or until the chain autocorrelation time estimate was changing by less than 1\% and the chain was longer than 100 times the autocorrelation time, whichever was shorter. The large number of model fits that must be run necessitates the upper limit on the number of steps, in order to complete the modelling in a reasonable time frame. We implemented a burn-in of 50\%, discarding the first half of the total number of chain steps, i.e. 10,000 steps discarded if the MCMC has run for 20,000 steps. This large burn-in ensures that the results are not biased by the initial conditions. We kept only those sub-bands for which the chains converged onto a single set of parameters for the model. Cases for which the chains failed to converge, or converged on two or more sets of parameters simultaneously, we concluded to be unsuccessful and excluded from the subsequent analysis. We calculated the best fit parameters to be the 50th percentiles of the MCMC samples, with the errors taken as the average of the differences between the 16th and 50th, and 50th and 84th percentiles respectively. We made use of GNU Parallel \citep{Tange2011} to run our modelling on multiple pulsars simultaneously. 
Given the values of $\tau$ obtained through our MCMC fit, we calculated $\alpha$. Initial calculations of $\alpha$ were done with least squares fitting of a power law to $\tau$ vs frequency. However, since it was found that different least squares algorithms tended to produce different answers, we decided instead to use a further MCMC to map out the probability distribution and gain a better understanding of convergence on the solution. We kept only those 142 pulsars which had at least 4 channels for which the $\tau$ fit was successful, and then also converged on a successful fit for $\alpha$. The value of $\alpha$ and its uncertainty we took from the same percentiles of the MCMC samples as described above.

\subsection{Applicability of methodology}
\label{sec:resultsfilter}

When drawing conclusions on the properties of a population of scattered pulsars, we require reliable measurements of the scattering timescale as a function of frequency. 
The first test is to check the diagnostic output of our methodology showing $\tau$ as a function of frequency and reject all 21 sources which showed no significant evolution of $\tau$ across the band. Even though some pulsars matched the selection criterion of a Gaussian profile convolved with an exponential, they showed no frequency evolution consistent with scattering. There is an expectation that time domain fits will break down when $\tau$ is too small to measure. If the underlying profile is not Gaussian in shape, this too will affect the fitting, particularly when $\tau$ is small. In Appendix \ref{app:theory} we show that attempting measurements of $\tau$ when $\sigma/\tau \geq 1$ is likely to lead to inaccuracy, where $\sigma$ is the standard deviation of the Gaussian describing the intrinsic pulse profile. The second test therefore is to inspect our results for $\tau$ and $\sigma$ and reject 31 sources with unreliable measurements. We describe how we identify this unreliability with an example in section \ref{sec:covar}. 
A further two pulsars (J1320--3512 and J1738--3211) we discarded due to the model being visibly inaccurate at the leading edge of the profile. The rising edges of the model and profile respectively did not overlap for these profiles. PSR~J1738--3211 has an asymmetric profile with the trailing edge steeper than the rising edge, implying that it is not scattered and should be discarded regardless. PSR~J1320--3512 has a leading edge that is shallower than the model leading edge. This suggests either that the intrinsic profile shape is not well modelled by a Gaussian, or that the scattering function for this pulsar might be better described by a thick screen model \citep{Williamson1972, Kirsten2019}, which has a more gradual leading edge shape than the thin screen model applied in this work. We rejected PSR~J1655--3844 after closer inspection at high frequencies revealed it not to fit our single-component criterion.
Of the pulsars remaining, there are 15 for which we have rejected a section of the band, for reasons detailed in section \ref{sec:nonPL} and Appendix \ref{app:sim}. The selection criterion of requiring 4 successful fits for $\tau$ then removes a further 3 pulsars from the sample.
Having completed the processing of the modelling output as described above, from our original sample of 205 pulsars we have a final subset of 84 for which we have scattering fits in which we can be confident, with at least 4 measured values of $\tau$.

\section{Results}
\label{sec:results}

Before presenting the results for our full set of pulsars, we show an example of a successful fit, and cases where the results of the fit are unreliable. 

\subsection{Modelling results: example PSR~J1818--1422}

\begin{figure*}
    \includegraphics[height=0.8\textheight,keepaspectratio]{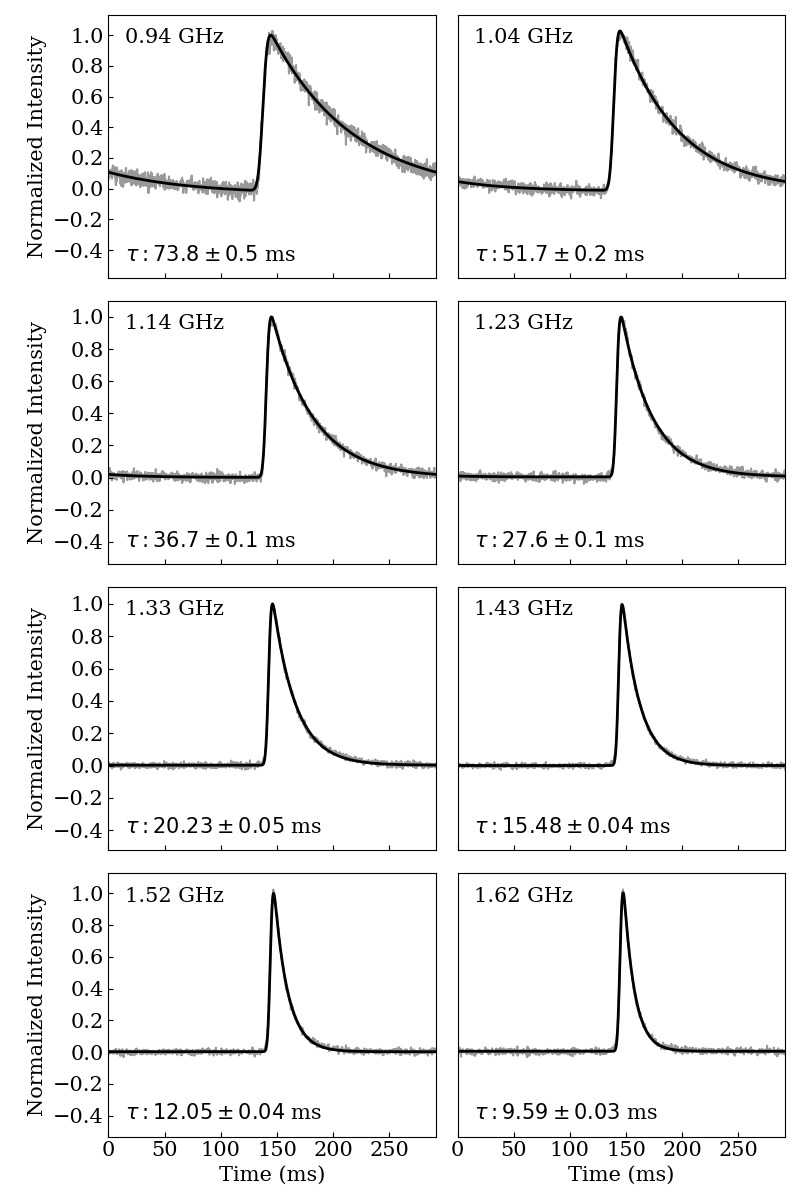} 
    \caption{Profiles of PSR J1818--1422 at 8 frequency channels across the observing band, plotted in grey, with the best fit scattering models over-plotted in black. The corrected frequency $f_{m}$ and scattering timescale, $\tau$, are shown on each profile. The profiles are ordered by increasing frequency, reading from left to right and top to bottom. }
    \label{fig:example_profs}
\end{figure*}

\begin{figure*}
    \includegraphics[width=\textwidth,height=\textheight,keepaspectratio]{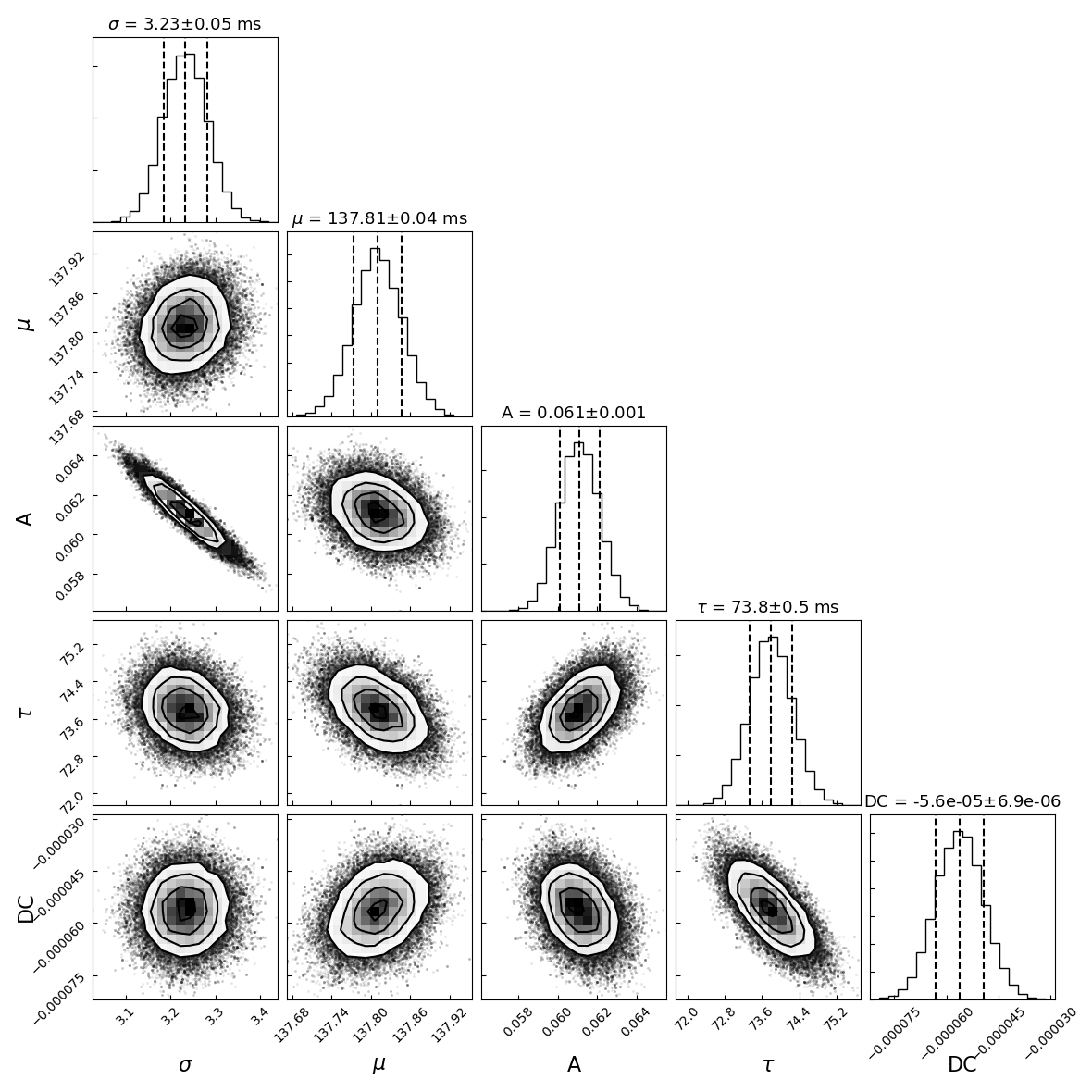}
    \caption{Corner plot for the scatter modelling parameters of the lowest frequency (950~MHz) profile of PSR J1818--1422. In each 2D subplot the black points indicate all of the sample values explored by the chains. On top of these points is plotted a 2D histogram and a contour plot to indicate the overall density distribution. The five model parameters are the standard deviation $\sigma$, mean $\mu$ and amplitude $A$ of the Gaussian, the scattering timescale $\tau$ and any DC offset of the pulse baseline from 0. We mark the 16th, 50th and 84th percentiles on the marginalized histograms. }
    \label{fig:example_corner}
\end{figure*}

\begin{figure}
    \includegraphics[width=\columnwidth]{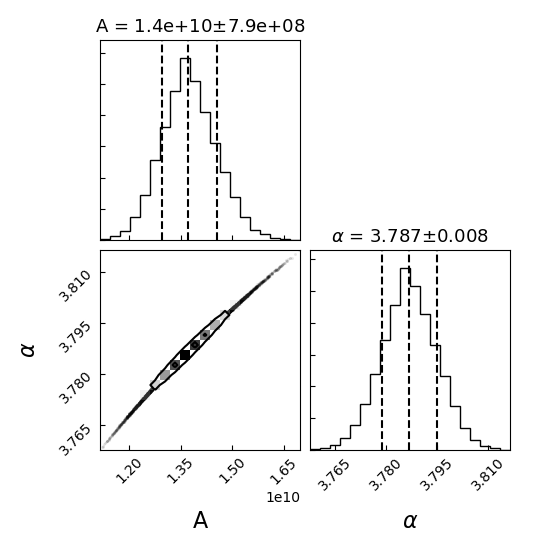}
    \caption{Corner plot for the power law fit parameters for scattering timescale $\tau$ against frequency $\nu$, where $\tau = A\nu^{-\alpha}$, $\alpha$ is the spectral index we wish to measure and $A$ is the constant of proportionality. The corner plot is formatted in the same way as that in Fig. \ref{fig:example_corner}. }
    \label{fig:alphacorner}
\end{figure}

\begin{figure}
    \includegraphics[width=\columnwidth]{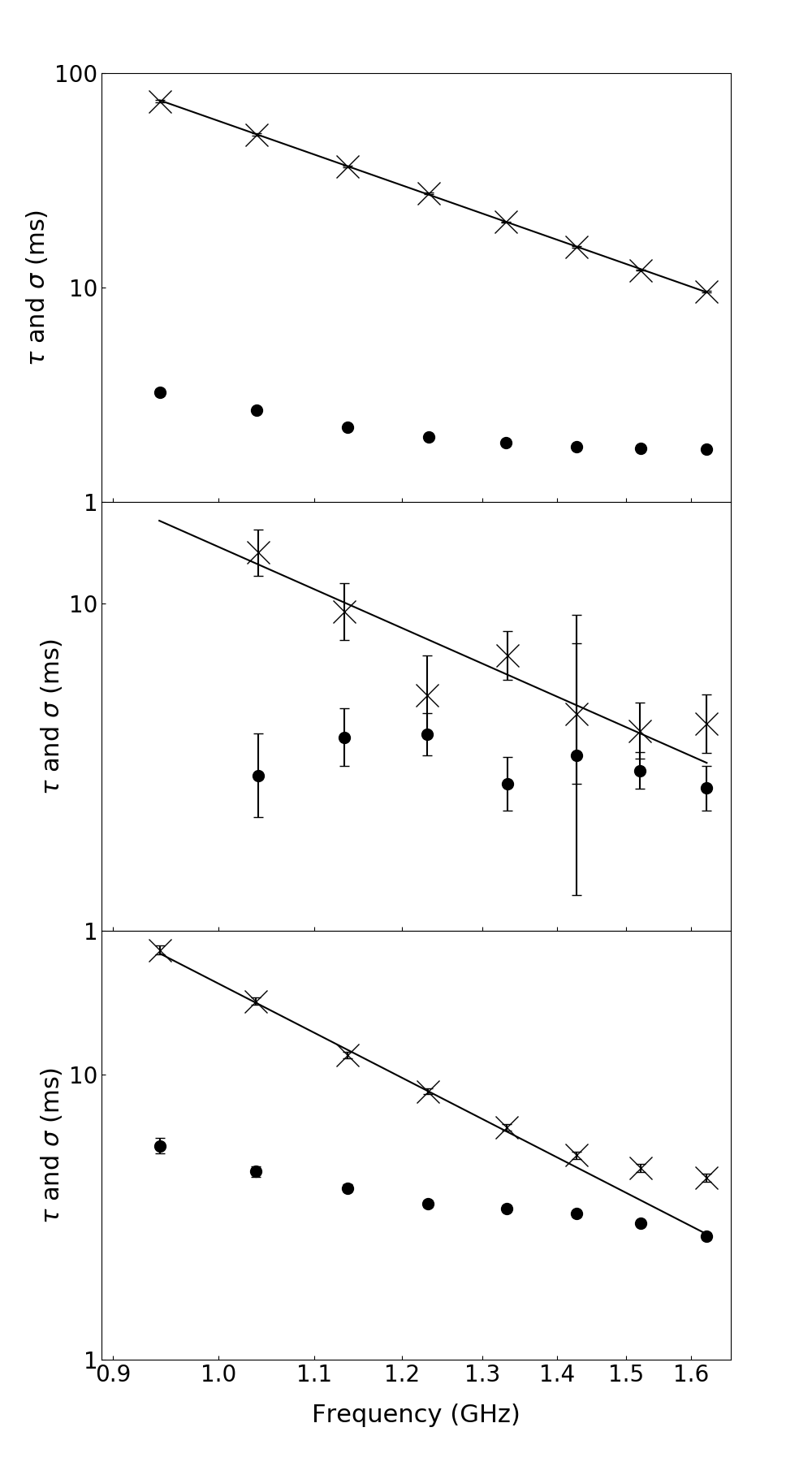}
    \caption{Log-log plots of scattering timescale $\tau$ (crosses) and intrinsic Gaussian standard deviation $\sigma$ (dots) against frequency for three pulsars, with the best fit power law plotted as a black line. Top: PSR~J1818--1422, an example of successful modelling. Middle: PSR~J1743--3153, an example of $\sigma-\tau$ mirroring; this pulsar and 32 similar sources are rejected from the final sample. Bottom: PSR~J1653--4249, an example of inaccurate modelling at high frequencies. For this pulsar we use only the lowest 5 frequencies to calculate the power law relationship; similar rejection of high frequency model results is done for a further 14 pulsars. }
    \label{fig:combi_tausigfreq}
\end{figure}

Fig. \ref{fig:example_profs} shows the eight observed pulse profiles of PSR~J1818--1422 with the best fit model at each frequency overlaid. We show how the parameters for the model at 0.94~GHz were obtained by representing the MCMC chain as a corner plot in Fig. \ref{fig:example_corner}. Fig. \ref{fig:alphacorner} shows the corner plot for the power law fit of $\tau = A\nu^{-\alpha}$, which gives $\alpha = 3.787 \pm 0.008$. The figure shows that $A$ and $\alpha$ are covariant, but that $\alpha$ is nevertheless tightly constrained. The top subplot of Fig. \ref{fig:combi_tausigfreq} shows how both $\tau$ and 
$\sigma$ evolve with frequency for this pulsar, with the power law fit indicated with a straight line. In addition, the modelling indicates intrinsic profile evolution, as $\sigma$ increases with decreasing frequency below $\sim$1~GHz, evolution comparable to that described for many other pulsar observations \citep[see for example][]{Thorsett1991}. 

\subsection{Covariant model parameters: an example}
\label{sec:covar}

We find that the results for several pulsars, particularly those that are less strongly scattered, exhibit an inter-relation between the measured values of $\sigma$ and $\tau$ that is not inherent to the pulsar and must result from the fitting process. An example is PSR~J1743--3153, for which we show a plot of both $\tau$ and $\sigma$ against frequency in the middle subplot of Fig. \ref{fig:combi_tausigfreq}. 
The decrease in $\tau$ with frequency seen for the first three measurements is mirrored by an increase in $\sigma$. The converse is then true for the next measurement: a drop in $\sigma$ is counteracted by an increase in $\tau$. Similar mirroring behaviour can also be seen in the measurements for the highest frequencies. This suggests that the modelling process has a covariance in $\sigma$ and $\tau$: multiple different pairings of these parameters all generate similarly shaped profiles and are therefore indistinguishable. Each sub-band is modelled individually, and the inaccuracy of model parameters converged upon for a given sub-band is then only apparent when compared to the results for the other sub-bands. 

It seems likely that PSR~J1743--3153 is scattered, but the mirroring behaviour of $\sigma$ and $\tau$ means we cannot trust the accuracy of the parameter measurements. We therefore exclude this pulsar, and 32 more sources showing similar behaviour, as identified by visual inspection of similar figures, from further analysis.

\subsection{Power law deviations and non-Gaussian pulse shapes}
\label{sec:nonPL}

There are 15 pulsars which show a visible deviation from a single power law at the higher frequencies, with $\tau$ larger than expected. 
We hypothesize that this may be caused by the intrinsic profile, and not the scatter broadening, being the dominant factor in the overall pulse shape at high frequencies for these pulsars. As shown in Appendix \ref{app:theory}, time domain fitting is unlikely to be accurate when $\sigma/\tau \geq 1$ and the intrinsic profile dominates over the scattering timescale. It is logical that there is a subset of pulsars for which scattering is measurable at lower frequencies, but the pulse shape is dominated by the intrinsic profile at higher frequencies. If so, our assumption of a single Gaussian component is insufficient to replicate the profile shape, resulting in over-estimates of $\tau$. We address as an example the case of PSR J1653--4249, for which $\tau$ and $\sigma$ are plotted against frequency in the bottom subplot of Fig. \ref{fig:combi_tausigfreq}. 

It is possible to simulate this pulsar using a profile with an additional weak trailing edge component. At high frequencies the absence of scattering is compensated by the presence of the trailing edge component, such that the measured values of $\tau$ do not continue to decrease with increasing frequency (see Appendix \ref{app:sim} for details). Although the simulation is very specific, it demonstrates that it is possible to obtain such a result under these conditions and therefore attributing the anomalous behaviour seen in the bottom subplot of Fig. \ref{fig:combi_tausigfreq} to properties of the ISM requires caution. 

In general, for this sample of 15 pulsars, we conclude that the intrinsic profile shape is the dominant factor in determining pulse shape at higher frequencies,  rather than the scattering timescale, and that as a result any non-Gaussianity in the intrinsic profile shape may be mis-modelled as contributing to the scattering tail. A possible alternative method for modelling pulse profiles with intrinsic shapes more complex than a Gaussian would be to apply the {\sc clean} algorithm. However, this would require its own set of assumptions and would result in modelling results that are not directly comparable to the rest of the results presented in this paper. For these 15 pulsars therefore, we kept only that frequency range where $\tau$ follows a power law, selecting the values through visual inspection, and inferred $\alpha$ from those.

\subsection{Scattering parameters for 84 pulsars}
\label{sec:final}

We present the results of the modelling in Table \ref{tab:results}. For each pulsar we list the scattering timescale $\tau$ at 1 GHz, along with $\alpha$ and the corrected dispersion measure (DM) that we measure through our calculation of the position of the intrinsic pulse profile at each frequency. We also indicate which of the 8 frequencies have been used to calculate these values, indicating which channels were excluded due either to failing to reach convergence on a solution, or to deviating from a power law at high frequencies, as explained in section \ref{sec:nonPL}. We use our MCMC power law fit $\tau = A\nu^{-\alpha}$ to compute $\tau$ at $\nu = 1$~GHz. We do this by calculating the value of $\tau$ at 1~GHz associated with every pair of values of $A$ and $\alpha$ explored by the MCMC chain, and taking the 50th quantile of these values as $\tau$ and the difference between the 16th and 84th quantiles for its error. We discuss the origin of our corrections to the DMs of these scattered pulsars, along with the relevance of using these corrected DMs, in section \ref{sec:DM}.

\begin{table}
\caption{Values for $\tau$ at 1~GHz, $\alpha$ and DM for our filtered sample of scattered pulsars. The values and their uncertainties are calculated as described in the text. The DM is the value that best aligns the modelled intrinsic profile at each frequency. We indicate which of the 8 sub-bands (ordered in increasing frequency) were used to generate these values with ones, and those not included with zeros. The minimum allowed number of sub-bands is 4. Reasons for exclusion of sub-bands (failure of model convergence or deliberate exclusion) are explained in the text.}
\label{tab:results}
\begin{tabular}{p{0.17\columnwidth}p{0.12\columnwidth}p{0.15\columnwidth}p{0.15\columnwidth}p{0.15\columnwidth}}
 \hline
\textbf{PSRJ} & \textbf{Which} & \textbf{$\tau$ at 1~GHz} & \textbf{$\alpha$} & \textbf{DM} \\
 & \textbf{channels} & \textbf{(ms)} & & \textbf{(cm$^{-3}$pc)} \\
 \hline
J0646+0905&11110110&10.03$\pm$0.03&3.46$\pm$0.05&147.85$\pm$0.07 \\
J1055$-$6028&11110000&2.36$\pm$0.08&4.5$\pm$0.4&636.9$\pm$0.1 \\
J1112$-$6103&00111111&33$\pm$3&3.8$\pm$0.2&595.69$\pm$0.09 \\
J1114$-$6100&11101101&28.3$\pm$0.1&3.14$\pm$0.02&676.7$\pm$0.1 \\
J1138$-$6207&11111111&26$\pm$2&3.8$\pm$0.2&518.86$\pm$0.05 \\
J1305$-$6203&11110000&10.4$\pm$0.2&3.1$\pm$0.2&468.78$\pm$0.05 \\
J1316$-$6232&00001111&1099$\pm$25&4.435$\pm$0.003&966.4$\pm$0.3 \\
J1319$-$6105&11110000&7.05$\pm$0.08&3.5$\pm$0.1&440.6$\pm$0.1 \\
J1341$-$6220&11111111&23.5$\pm$0.2&4.04$\pm$0.02&718.26$\pm$0.06 \\
J1349$-$6130&11111100&6.4$\pm$0.1&3.9$\pm$0.1&283.87$\pm$0.04 \\
J1406$-$6121&00011111&48$\pm$1&4.751$\pm$0.004&537.29$\pm$0.08 \\
J1412$-$6145&11111111&19.9$\pm$0.5&4.0$\pm$0.1&512.48$\pm$0.08 \\
J1413$-$6141&11111111&42$\pm$2&4.1$\pm$0.1&667.60$\pm$0.06 \\
J1511$-$5835&11111100&23.3$\pm$0.5&4.8$\pm$0.2&329.4$\pm$0.2 \\
J1512$-$5759&11111111&7.14$\pm$0.02&3.16$\pm$0.01&626.90$\pm$0.02 \\
J1514$-$5925&00111111&16.7$\pm$0.6&5.107$\pm$0.005&192.51$\pm$0.08 \\
J1519$-$5734&01011101&118$\pm$7&3.9$\pm$0.2&654.7$\pm$0.4 \\
J1538$-$5551&11111111&22$\pm$2&4.0$\pm$0.2&602.67$\pm$0.06 \\
J1543$-$5459&11111111&36.2$\pm$0.8&3.76$\pm$0.08&344.99$\pm$0.03 \\
J1551$-$5310&11111111&131$\pm$5&4.4$\pm$0.1&485.8$\pm$0.6 \\
J1610$-$5006&11111000&124$\pm$1&4.670$\pm$0.002&410.0$\pm$0.2 \\
J1630$-$4719&11111110&5.71$\pm$0.09&3.3$\pm$0.2&487.2$\pm$0.1 \\
J1630$-$4733&11110000&340$\pm$2&5.049$\pm$0.001&509.3$\pm$0.2 \\
J1632$-$4621&11111110&16.85$\pm$0.07&3.87$\pm$0.03&559.83$\pm$0.07 \\
J1633$-$4453&11111111&19.59$\pm$0.08&3.99$\pm$0.02&472.22$\pm$0.03 \\
J1638$-$4608&01111011&17$\pm$1&4.6$\pm$0.2&422.45$\pm$0.10 \\
J1640$-$4715&11111111&56$\pm$1&4.45$\pm$0.08&581.0$\pm$0.2 \\
J1640$-$4951&11111111&14.4$\pm$0.6&4.5$\pm$0.3&407.4$\pm$0.2 \\
J1650$-$4341&00011111&66$\pm$7&4.4$\pm$0.3&672.8$\pm$0.4 \\
J1653$-$4249&11111000&20.9$\pm$0.2&4.17$\pm$0.05&415.36$\pm$0.06 \\
J1700$-$4422&11101100&75$\pm$4&4.7$\pm$0.5&404.9$\pm$0.9 \\
J1702$-$4128&11111111&25.5$\pm$0.7&3.97$\pm$0.08&365.75$\pm$0.08 \\
J1707$-$4053&11111011&95.2$\pm$0.2&3.950$\pm$0.007&351.78$\pm$0.08 \\
J1715$-$3859&11111111&228$\pm$9&4.1$\pm$0.1&806.2$\pm$0.6 \\
J1717$-$3425&11011010&19.82$\pm$0.02&3.466$\pm$0.007&583.46$\pm$0.01 \\
J1717$-$3737&11111111&35.7$\pm$0.5&3.42$\pm$0.06&522.70$\pm$0.05 \\
J1719$-$4006&11101000&5.10$\pm$0.09&3.4$\pm$0.2&386.12$\pm$0.04 \\
J1720$-$3659&11111111&12.6$\pm$0.2&3.63$\pm$0.07&378.97$\pm$0.03 \\
J1721$-$3532&11111111&113.4$\pm$0.7&4.10$\pm$0.02&493.0$\pm$0.1 \\
J1724$-$3149&11111000&40$\pm$2&4.928$\pm$0.007&400.8$\pm$0.3 \\
J1725$-$3546&11111110&50$\pm$2&4.1$\pm$0.1&738.3$\pm$0.4 \\
J1730$-$3350&11111111&21.0$\pm$0.4&4.00$\pm$0.05&260.40$\pm$0.04 \\
J1731$-$3123&11111111&15.1$\pm$0.5&1.3$\pm$0.1&354.5$\pm$0.2 \\
J1739$-$3131&11111111&68.2$\pm$0.2&3.836$\pm$0.009&596.9$\pm$0.1 \\
J1740$-$3052&11111000&12.1$\pm$0.1&3.83$\pm$0.10&738.80$\pm$0.05 \\
J1801$-$2304&01111000&547$\pm$10&4.529$\pm$0.003&1067.85$\pm$0.10 \\
J1811$-$1736&01111111&42$\pm$1&3.24$\pm$0.09&473.93$\pm$0.04 \\
J1812$-$1718&11111101&31.8$\pm$0.2&3.95$\pm$0.03&251.4$\pm$0.1 \\
J1812$-$1733&10101111&102$\pm$1&3.40$\pm$0.04&509.8$\pm$0.1 \\
J1816$-$1729&11111000&14.3$\pm$0.1&4.47$\pm$0.07&520.7$\pm$0.2 \\
J1818$-$1422&11111111&59.8$\pm$0.2&3.787$\pm$0.008&619.65$\pm$0.07 \\
J1818$-$1607&11111100&45.0$\pm$0.5&4.41$\pm$0.08&699.2$\pm$0.8 \\
J1819$-$1114&11111111&51$\pm$2&4.4$\pm$0.1&309.7$\pm$0.2 \\
J1819$-$1510&11111111&19.7$\pm$0.4&3.99$\pm$0.08&418.14$\pm$0.04 \\
J1820$-$1346&11111111&110.9$\pm$0.5&3.81$\pm$0.01&771.0$\pm$0.2 \\
\hline
\end{tabular}
\end{table}

\begin{table}
\contcaption{}
\label{tab:results2}
\begin{tabular}{p{0.17\columnwidth}p{0.12\columnwidth}p{0.15\columnwidth}p{0.15\columnwidth}p{0.15\columnwidth}}
 \hline
\textbf{PSRJ} & \textbf{Which} & \textbf{$\tau$ at 1~GHz} & \textbf{$\alpha$} & \textbf{DM} \\
 & \textbf{channels} & \textbf{(ms)} & & \textbf{(cm$^{-3}$pc)} \\
 \hline
J1822$-$1400&11111111&6.81$\pm$0.09&3.31$\pm$0.08&649.27$\pm$0.04 \\
J1824$-$1118&11111111&26.9$\pm$0.2&3.77$\pm$0.02&601.31$\pm$0.06 \\
J1824$-$1159&11111111&18.8$\pm$0.3&3.28$\pm$0.06&462.96$\pm$0.09 \\
J1824$-$1423&11111111&12.8$\pm$0.3&4.2$\pm$0.1&427.63$\pm$0.08 \\
J1825$-$1446&11111110&21$\pm$1&4.0$\pm$0.2&351.6$\pm$0.1 \\
J1832$-$1021&11111111&12.49$\pm$0.07&4.05$\pm$0.03&474.14$\pm$0.03 \\
J1833$-$0559&11111111&180$\pm$6&3.78$\pm$0.09&346.70$\pm$0.10 \\
J1834$-$0731&00101111&122$\pm$9&3.9$\pm$0.2&288.3$\pm$0.4 \\
J1835$-$0643&11111111&110$\pm$3&3.77$\pm$0.07&464.8$\pm$0.1 \\
J1837$-$0604&00011101&62$\pm$6&5.01$\pm$0.02&462$\pm$10 \\
J1839$-$0321&00111111&14$\pm$2&5.0$\pm$0.4&450.5$\pm$0.1 \\
J1839$-$0643&11111111&47.3$\pm$0.7&4.19$\pm$0.05&493.5$\pm$0.2 \\
J1840$-$0559&11111111&23.5$\pm$0.4&3.85$\pm$0.09&319.1$\pm$0.1 \\
J1841$-$0425&11111110&3.22$\pm$0.03&3.29$\pm$0.07&324.77$\pm$0.02 \\
J1842$-$0153&11111111&32.7$\pm$0.3&3.92$\pm$0.05&422.9$\pm$0.1 \\
J1844$-$0030&11111000&12.3$\pm$0.3&4.1$\pm$0.2&603.2$\pm$0.1 \\
J1844$-$0244&11101110&20.0$\pm$0.4&3.31$\pm$0.09&422.13$\pm$0.06 \\
J1844$-$0538&11111100&11.73$\pm$0.07&4.07$\pm$0.03&410.51$\pm$0.04 \\
J1846$-$0749&11111100&7.14$\pm$0.06&3.62$\pm$0.06&389.13$\pm$0.02 \\
J1850$-$0006&11111111&260$\pm$7&4.2$\pm$0.1&625$\pm$2 \\
J1850$-$0026&11111000&46.5$\pm$0.6&4.946$\pm$0.002&948.8$\pm$0.2 \\
J1852$-$0127&01111111&57$\pm$3&3.6$\pm$0.1&427.9$\pm$0.2 \\
J1853+0545&11111111&20.0$\pm$0.3&3.13$\pm$0.04&197.91$\pm$0.03 \\
J1857+0143&11111111&50$\pm$2&4.0$\pm$0.1&247.9$\pm$0.1 \\
J1857+0526&11111111&24.0$\pm$0.6&3.78$\pm$0.07&464.79$\pm$0.05 \\
J1859+0601&00011111&125$\pm$26&4.4$\pm$0.7&272.4$\pm$0.5 \\
J1913+1145&01111111&15.2$\pm$0.7&3.9$\pm$0.2&642.0$\pm$0.2 \\
J1916+0844&11111100&12.8$\pm$0.1&4.14$\pm$0.07&338.01$\pm$0.07 \\
J1928+1923&11111110&48.6$\pm$0.5&3.94$\pm$0.05&476.4$\pm$0.2 \\
\hline
\end{tabular}
\end{table}

\section{Discussion}
\label{sec:disc}

\subsection{The scattering spectral index distribution}
\label{sec:alphadisc}

\begin{figure}
    \includegraphics[width=\columnwidth]{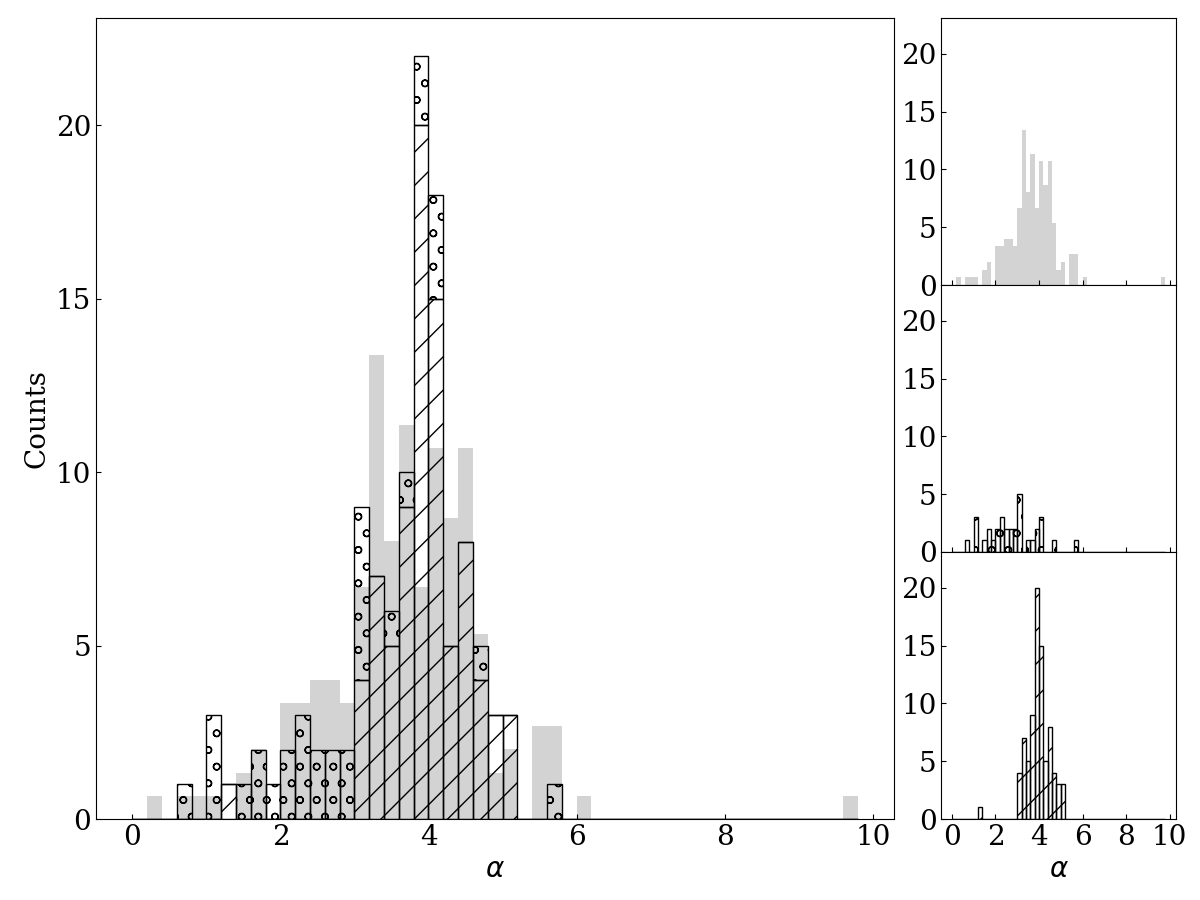}
    \caption{Histograms of values of $\alpha$. Top right (grey): previously published values of $\alpha$ obtained through time domain scatter modelling. Middle right (transparent, circle pattern): values of $\alpha$ for the pulsars rejected from our sample due to evidence of inaccurate modelling. Bottom right (transparent, hatched): histogram of $\alpha$ values for our successful sample of 84 pulsars. Left: the same three histograms, now overlaid (and, for our observations, stacked) for visual comparison. }
    \label{fig:alphahist}
\end{figure}

Fig. \ref{fig:alphahist} shows histograms of the distributions of $\alpha$ obtained for our survey. We show the final filtered sample of 84 pulsars (hatched histogram) and the 33 pulsars which were filtered out due to problems in the modelling (histogram with circle pattern). These have primarily been removed due to covariance between the values of $\tau$ and $\sigma$, as described in sections \ref{sec:resultsfilter}, \ref{sec:covar} and appendix \ref{app:theory}, but this subset also includes the two pulsars rejected due to the leading edge not being correctly modelled (PSRs J1320--3512 and J1738--3211). The removal of pulsars is done without reference to the value of $\alpha$ obtained, so that we are not biased in favour of those pulsars with values of $\alpha$ close to what might be expected from theory. The result of the filtering is that pulsars with small values of $\alpha$ are largely removed from the sample, and those remaining show a near-symmetric distribution. We have shown, in Section \ref{sec:nonPL} and in the appendices, that poor model fits have a tendency to overestimate $\tau$, and that poor model fits are more likely for smaller vales of $\tau$, which are seen more often at higher frequencies. It is therefore expected that poorly modelled pulsars will tend to underestimate $\alpha$, and so it is unsurprising that the majority of the poorly modelled pulsars returned small values of $\alpha$.

Fig. \ref{fig:alphahist} also shows a histogram of the values of $\alpha$ found in the literature for cases where pulsar profiles observed at $\geq 400$~MHz have been fitted with an isotropic time domain model \citep{Cordes1985, Lohmer2001, Lohmer2004, Lewandowski2011, Lewandowski2013, Lewandowski2015, Lewandowski2015a, Krishnakumar2017, Krishnakumar2019}. The histogram of literature values has been weighted so that it represents the same total count of pulsars as those shown for this work. We apply the $\geq400$~MHz cut because we observe different behaviour in the scattering surveys at lower frequencies, which we discuss in section \ref{sec:complit}.

\begin{figure}
    \includegraphics[width=\columnwidth]{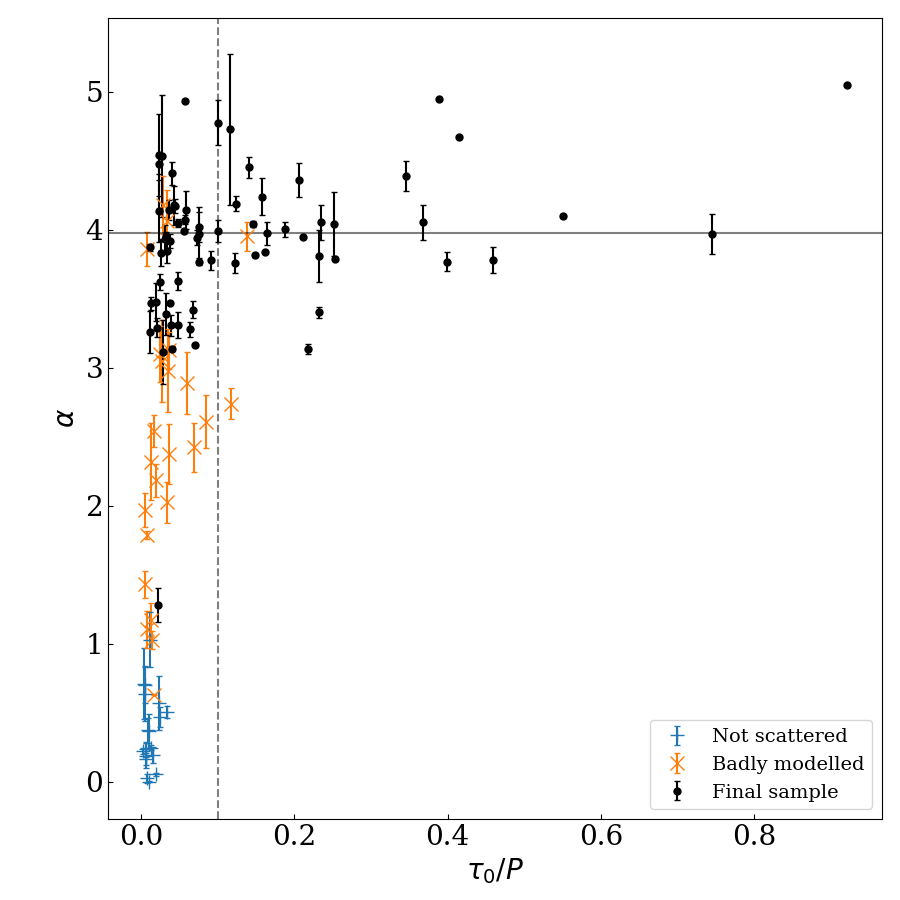}
    \caption{Plot of scattering spectral index $\alpha$ against scattering timescale at the lowest frequency channel of 950~MHz, $\tau_{0}$, where $\tau_{0}$ is scaled by pulse period $P$. Blue vertical crosses: non-scattered pulsars rejected from the final sample. Orange diagonal crosses: pulsars rejected due to evidence of inaccurate modelling. Black points: final filtered sample of successfully modelled scattered pulsars.}
    \label{fig:alphatau0}
\end{figure}

Comparing our distribution of $\alpha$ with that of previously published values, we see that, when we include our poorly modelled pulsars, the distributions are similar. The population of pulsars around $0 < \alpha < 3$ shown in the published values has been filtered out from our own distribution. Performing a Kolmogorov-Smirnov (KS) test to compare our sample to that from the literature, we obtain the following values. When we use just our final filtered sample in the KS test, we obtain a KS statistic of 0.247 with a corresponding p-value of 0.001. Adding in our poorly modelled pulsars as well gives KS statistic 0.109 and p-value 0.335. This indicates that, whereas our final filtered sample is clearly not drawn from the same distribution as that of the combined literature values, when we include our poorly modelled pulsars in our distribution, we cannot reject the null hypothesis. 
Fitting a Gaussian to our filtered distribution of $\alpha$ we find that it has a mean of 4.0 and a standard deviation of 0.6. The theoretical values of both $\alpha = 4$ and $\alpha = 4.4$ fall within one standard deviation. Fig. \ref{fig:alphatau0} shows our measurements of $\alpha$ plotted against $\tau_{0}/P$, where $\tau_{0}$ is $\tau$ measured for our lowest observing frequency of 950~MHz. It is interesting to note that if we split our population into sources where $\tau_{0}$ is less or greater than 10\% of the period, we find a mean $\alpha$ value for each group of $3.7 \pm 0.6$ and $4.1 \pm 0.4$ respectively. For $\alpha = 4$, a pulsar with $\tau_{0}/P = 0.1$ at the lowest frequency will have $\tau/P \sim 0.01$ at the highest frequency. We have shown in section \ref{sec:nonPL} and in the appendices that our methodology overestimates small values of $\tau$.
This suggests that we measure systematically smaller values of $\alpha$ for such cases, which is a limitation of the time domain methodology. 
Fig. \ref{fig:alphatau0} shows that the sources for which $\tau_{0}/P < 0.1$ are comprised of pulsars that we have filtered out as being either poorly modelled (orange diagonal crosses), or not scattered (blue vertical crosses), or they are not filtered out (black points) and yet still show systematically lower values of $\alpha$, as described above. 

We scaled the scattering timescale by the period for the analysis above because all of our observations have the same number of bins across the pulse period (1024 bins). The number of bins containing information about the profile shape will affect the accuracy of the modelling, and a profile that takes up a smaller fraction of the pulse period (smaller $\tau/P$) will have fewer bins spanning the profile. It is therefore the ratio $\tau/P < 0.1$ that is important in terms of likelihood of the model being able to correctly capture the scattering behaviour of a given pulsar. However, the time resolution of the bins is also relevant to consider. The larger a pulsar's period, the lower the time resolution of a single bin, meaning that less information is captured in each bin. We find that the distribution of periods for the $\tau/P < 0.1$ cases is shifted slightly towards longer periods than the distribution for the $\tau/P > 0.1$ cases, which reflects the time resolution limitation.

\subsection{Correcting for dispersion}
\label{sec:DM}

As part of our scatter modelling we identify the position of the underlying Gaussian that describes the intrinsic profile in our model. By fitting for a correction to the DM on these values, we obtain the DM that best aligns the intrinsic pulse profile independent of the effects of scattering. Our new values for the DMs of the pulsars in our sample are presented in Table \ref{tab:results}. An alternate approach would be to simultaneously model the data in all sub-bands, which would enable fitting for DM and the pulse scattering index.  This approach was taken in modelling the pulse profiles of fast radio bursts (FRBs) in \cite{Qiu2020a}.

\begin{figure}
    \includegraphics[width=\columnwidth]{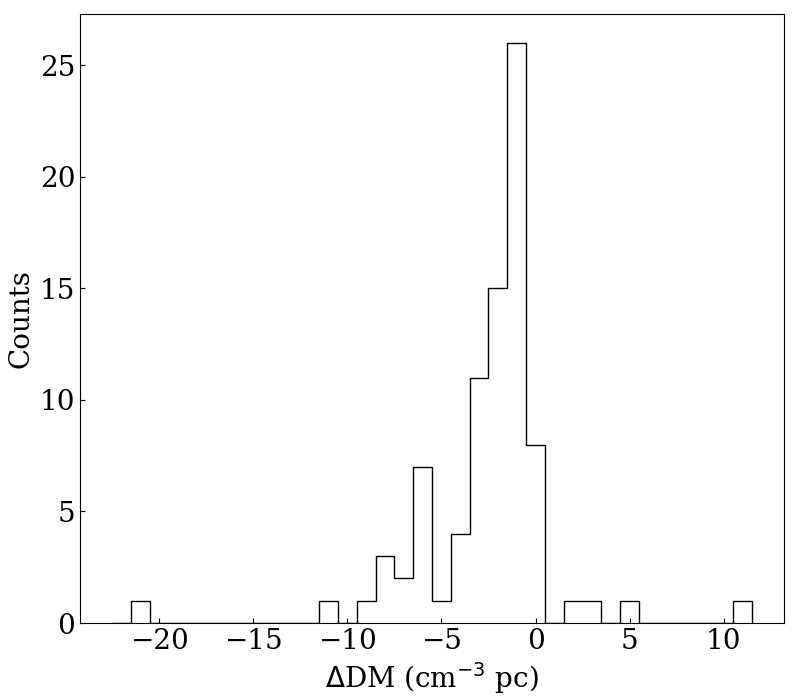}
    \caption{Histogram of the values of $\Delta$DM calculated for our filtered scattered pulsar sample, where $\Delta$DM is the difference between the original DM used to dedisperse the pulsar, and the best fit DM generated by our scattering model. }
    \label{fig:deltaDM}
\end{figure}

We show a histogram of the measured corrections to the DM in Fig. \ref{fig:deltaDM}. The majority of these are small and negative, as expected \citep[e.g.][]{Geyer2017a}. 
The tail of large negative $\Delta$DM values is partially attributable to the small subset of pulsars that are more strongly scattered and therefore require larger corrections to the DM to align intrinsic profiles. However, it also reflects the distribution of periods in our pulsar sample. This can be understood as follows. We can assume very roughly that pulsars have the same duty cycle. For profiles at two discrete frequencies that are misaligned by an amount on the order of the pulse width, the shift required to align the profiles will be the same fraction of the pulse period. For pulsars with very different periods, this shift will comprise different absolute lengths of time; correspondingly the DM correction to perform this shift will be larger for the pulsar with the larger period.

\begin{figure}
    \includegraphics[width=\columnwidth]{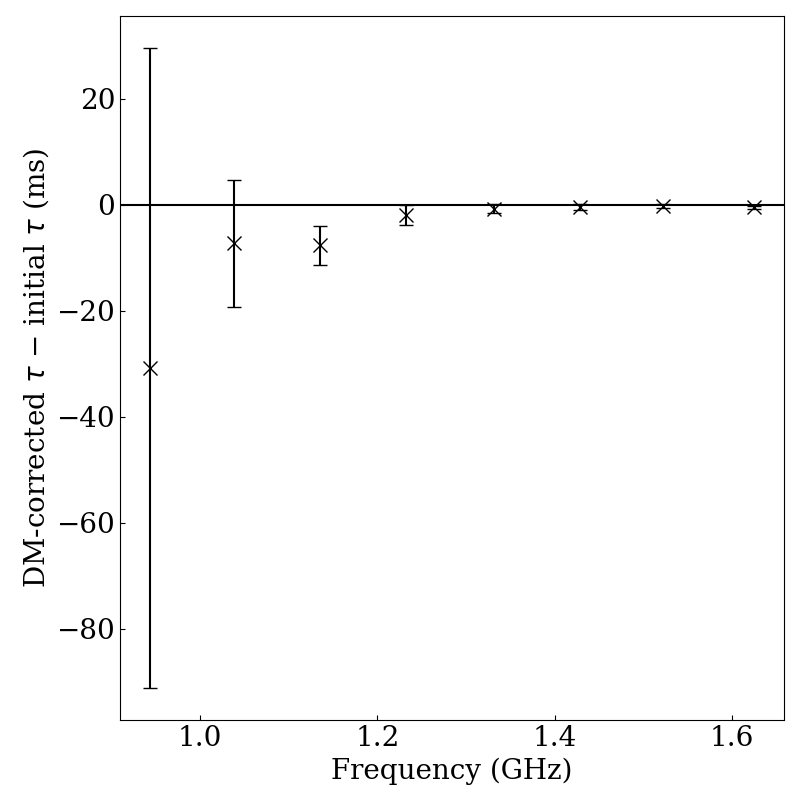}
    \caption{Plot of $\Delta \tau$ vs. frequency for PSR~J1630--4733, where $\Delta \tau$ is the difference between the scattering timescales calculated for the data processed with the initial, visibly incorrect, DM, and when reprocessed with the best fit DM produced by the scattering model.}
    \label{fig:compareDM}
\end{figure}

Since the choice of DM affects the profile shape within each sub-band, 
a question arises as to whether correcting the DM would change the measurements of $\tau$. 
We tested this on PSR~J1630--4733, chosen because its $\Delta$DM of 11.3 cm$^{-3}$pc results in the largest relative shift of profiles at different frequencies in our sample set. PSR~J1850--0006 has a larger absolute $\Delta$DM of $-$21.0 cm$^{-3}$pc, however, since its period is also larger, the corresponding time shift between profiles at different frequencies is a smaller fraction of the period and so the relative misalignment of profiles is smaller. We identified the magnitude of the DM correction required through scatter modelling, then re-processed the data with the new DM and then repeated the scatter modelling. We compare the results for $\tau$ obtained before and after the reprocessing in Fig. \ref{fig:compareDM}. The difference is largest at low frequencies, as expected. 
It is also uniformly negative, meaning that the scattering timescales in the uncorrected DM case are larger. However, even for this pulsar, where the change in DM is most extreme (shifting the top of the observing band with respect to the bottom of the band by 6\% of the pulse period), the measured values of $\tau$ are still equivalent to within 1$\sigma$ in all but the third channel (counting from lowest to highest frequency), and equivalent to within 3$\sigma$ for all channels.

\subsection{Scattering parameters in context}
\label{sec:complit}

\begin{table}
\caption{Table of pulsars for which there have been previous measurements of $\alpha$. We list the values of $\alpha$ measured in this work and those values found in the literature, with the appropriate citations. The reference codes in the table correspond to the following papers: \protect\citetalias{Lohmer2001} is \protect\cite{Lohmer2001}; \protect\citetalias{Lewandowski2013} is \protect\cite{Lewandowski2013}; \protect\citetalias{Lewandowski2015a} is \protect\cite{Lewandowski2015a}. }
\label{tab:compare}
\begin{tabular}{llll}
 \hline
\textbf{PSRJ} & \textbf{$\alpha$} & \textbf{$\alpha$} & \textbf{Reference} \\
 & \textbf{(this work)} & \textbf{(literature)} & \\
 \hline
J1801$-$2304 & 4.529$\pm$0.003 & 3.9$\pm$0.4 & \citetalias{Lohmer2001} \\
  & & 4.92$\pm$0.11 & \citetalias{Lewandowski2013} \\
  & & 3.45$\pm$0.09 & \citetalias{Lewandowski2015a} \\
J1818$-$1422 & 3.787$\pm$0.008 & 3.5$+$0.5$-$0.6 & \citetalias{Lohmer2001} \\
  & & 3.97$\pm$0.12 & \citetalias{Lewandowski2013} \\
J1820$-$1346 & 3.81$\pm$0.01 & 3.3$+$0.7$-$0.5 & \citetalias{Lohmer2001} \\
J1822$-$1400 & 3.31$\pm$0.08 & 4.1$+$0.7$-$1.2 & \citetalias{Lohmer2001} \\
  & & 3.96$\pm$0.18 & \citetalias{Lewandowski2013} \\
J1824$-$1118 & 3.77$\pm$0.02 & 3.5$+$0.7$-$0.6 & \citetalias{Lohmer2001} \\
J1825$-$1446 & 4.0$\pm$0.2 & 3.77$\pm$0.24 & \citetalias{Lewandowski2013} \\
J1835$-$0643 & 3.77$\pm$0.07 & 4.37$\pm$0.3 & \citetalias{Lewandowski2013} \\
J1841$-$0425 & 3.29$\pm$0.07 & 3.91$\pm$0.14 & \citetalias{Lewandowski2013} \\
J1857+0143 & 4.0$\pm$0.1 & 4.69 & \citetalias{Lewandowski2013} \\
\hline
\end{tabular}
\end{table}

Cross-comparing our sample of pulsars with the literature, we find 9 pulsars in our sample that have previous measurements for $\alpha$. We list these measurements in Table \ref{tab:compare}. Some of these have been measured multiple times, giving a total of 13 measurements of $\alpha$ to compare to our results, found in \cite{Lohmer2001}, \cite{Lewandowski2013} and \cite{Lewandowski2015a}. We note that 6 of the 7 estimates of $\alpha$ given in \cite{Lewandowski2013} are larger than our corresponding estimates, whilst 4 of the 5 estimates given in \cite{Lohmer2001} are smaller than ours. This suggests that differing modelling approaches may tend to give systematic differences in the resultant parameters estimated. Of the 9 pulsars, 5 have at least one measurement consistent with our own, to within the uncertainties. The 4 that are entirely inconsistent are all found in \cite{Lewandowski2013} and all have larger values of $\alpha$ quoted there than those presented here.

\begin{figure}
    \includegraphics[width=\columnwidth]{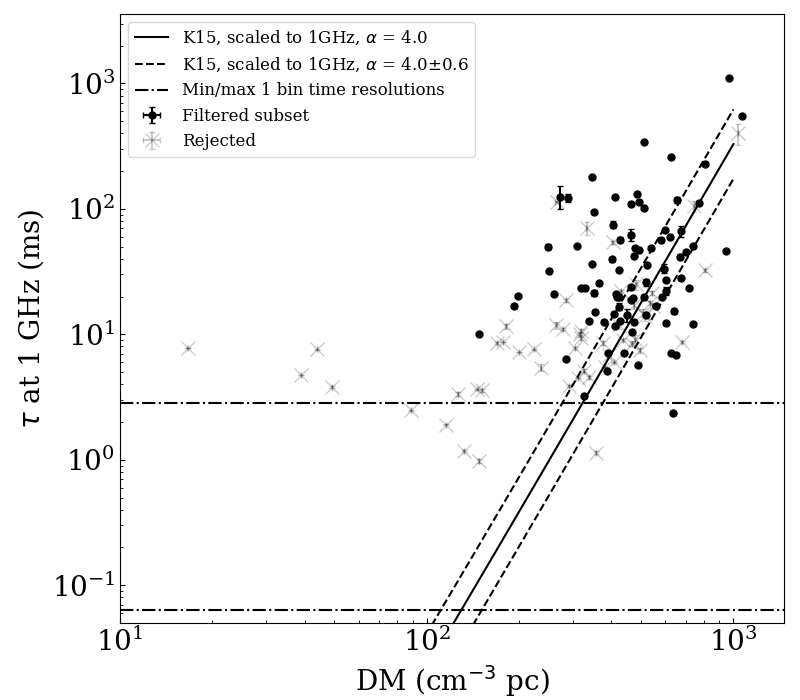}
    \caption{Plot of scattering timescale $\tau$ at 1~GHz vs. DM. Black points: the values for the final filtered sample. Black line: the best fit equation identified by \protect\cite{Krishnakumar2015}, scaled to 1~GHz using our best fit $\alpha$ of 4.0. Dashed lines: the same equation, now scaled using $\alpha = 4.0\pm0.6$. Pale grey crosses: pulsars rejected from the sample due either to inaccurate modelling or the pulsars not being scattered. Horizontal dot-dash lines: these mark the minimum and maximum time resolutions of a single bin for the pulsar sample, where all observations had a resolution of 1024 bins across the pulse period. }
    \label{fig:tau_DM_context}
\end{figure}

Fig. \ref{fig:tau_DM_context} shows $\tau$ calculated at 1~GHz plotted against DM. The values for the final sample we plot as black circles, and for comparison we also plot those values obtained from pulsars subsequently determined to be either not scattered or poorly modelled. The horizontal lines mark the smallest and largest time resolutions for a single bin of a pulse profile. Since all observations have 1024 bins across the profile, these values correspond to $1/1024^{\textrm{th}}$ of the smallest and largest periods in the sample. 
We take the equation relating $\tau$ and DM fit by \cite{Krishnakumar2015} at 327~MHz, scale it to 1~GHz using our best fit $\alpha$ of 4.0 and plot it as a black line. We also plot the scaled fits for $\pm\sigma$ about the mean, $\alpha = 3.4$ and $\alpha = 4.6$, as dashed lines. The functional form of this equation is
\begin{eqnarray}
\label{eq:tauDMKrishnakumar2015}
\tau_{\rm s} ~=~ 3.6 \times 10^{-6} \,\, {\rm DM}^{2.2}(1.0 + 0.00194 {\rm DM}^2)(1000/327)^{-4.0}, 
\end{eqnarray}
where $\tau$ has units of ms and DM has units cm$^{-3}$pc.
We see from Fig. \ref{fig:tau_DM_context} that the pulsars in our final sample have high DMs and that they are clustered close to the best fit line of \cite{Krishnakumar2015}. There is some spread in the values, meaning that our results are also compatible with previous fits in the literature made by \cite{Ramachandran1997, Lohmer2004, Bhat2004a} and \cite{Lewandowski2015}. Since our sample is limited to a narrow DM range, it is not possible to perform an equivalent model fit for the $\tau$-DM relationship using our own results. We note that our results are not symmetrically distributed about the \cite{Krishnakumar2015} model. This may be related to a difference in modelling approach: whereas we model the intrinsic pulse profile as a Gaussian without constraining its parameters, \cite{Krishnakumar2015} use a high frequency unscattered profile as the template for the intrinsic pulse profile. 

The choice of how to model the intrinsic pulse profile necessarily affects the results for the measured scattering timescales and spectral indices. For example, the slower rise time of the thick screen model may absorb the intrinsic pulse shape into its measurement, or conversely the choice of a Gaussian intrinsic pulse may obscure evidence for a thick screen scattering function. Another example is the comparison of our scattering timescale result for PSR~J1316--6232 with the estimate published by \cite{Crawford2001} of $\tau\sim150$~ms at 1.35~GHz. Scaling our result to this frequency gives a scattering timescale almost twice as large. This difference follows directly from the difference in modelling choices. Whereas \cite{Crawford2001} provide an estimate of the scattering timescale based only on the exponential decay of the profile intensity, our model involves a convolution of that exponential decay with a Gaussian representing the intrinsic profile shape. For this reason, we caution against direct comparison of individual $\tau$ values measured using different methods. A systematic shift of $\tau$ values caused by the choice of method is likely to have less of an impact on $\alpha$, provided all the values of $\tau$ used to calculate it were estimated using the same method. Nevertheless, this example highlights the importance of applying the same modelling approach to a large-scale sample of pulsars, such as this one, in order to be able to compare the scattering results for different pulsars. 

Of the values of $\tau$ associated with poor modelling, there are several at smaller DMs for which $\tau$ is larger than would be expected from the \cite{Krishnakumar2015} model. This is easily explained based on the results of our investigations into the causes of poor model fits. It is expected that pulsars with smaller DMs will, in general, be less strongly scattered. 
Our work shows that small scattering timescales are likely to be over-estimated, particularly when the timescale size is similar to the width of the intrinsic pulse profile. These poorly modelled pulsars also have $\tau$ estimates close to the temporal resolutions of the pulsars. Indeed, some of these pulsars were concluded not to be scattered at the observing band, which takes the concept of small values of $\tau$ being overestimated to the logical extreme. 
We caution therefore that there may be a wider tendency to overestimate $\tau$ at the lower limits of both DM and $\tau$ attainable at a given frequency.

\begin{figure}
    \includegraphics[width=\columnwidth]{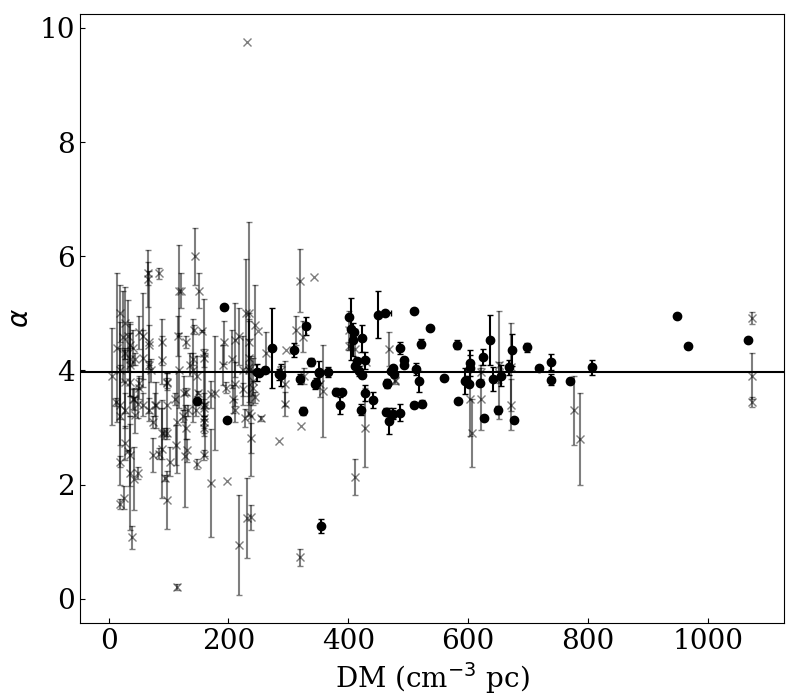}
    \caption{Plot of $\alpha$ vs. DM. Black points: our sample of 84 pulsars. Pale grey crosses: previously published values obtained through time domain scatter modelling. Horizontal line: the mean value of $\alpha$ calculated from a Gaussian fit to the distribution of our results.}
    \label{fig:alpha_DM_context}
\end{figure}

This may go some way to explaining what we observe in Fig. \ref{fig:alpha_DM_context}, which is a plot of $\alpha$ against DM for our values (black circles) and those given in the literature as described in section \ref{sec:alphadisc} and shown in Fig. \ref{fig:alphahist}. Our best fit of $\alpha = 4.0$, shown as a horizontal black line, is consistent with the results for the literature. On calculating mean $\alpha$ values for 5 bins across the DM range we saw no significant evolution of $\alpha$ with DM, save for a slight increase in $\alpha$ at the highest DM bin, encompassing the three points in Fig. \ref{fig:alpha_DM_context} at around 1000~cm$^{-3}$pc). More data is required to determine whether this increase is statistically significant. We also note that weighting the averages by the uncertainties in $\alpha$ tends to favour higher values in comparison to the unweighted averages. Inspecting the literature values of $\alpha$, we note a greater spread in the literature values at lower DMs. 
As we have described, the over-estimation of smaller values of $\tau$ tends to lead to under-estimation of $\alpha$. The reduced strength of scattering at lower DMs should cause this to have a greater effect at low DMs, which is what we see in Fig. \ref{fig:alpha_DM_context}.

There is a further consideration to which attention must be brought: whereas our modelling explains the distributions of $\alpha$ and $\tau$ obtained for pulsar observations at frequencies $\geq~400$~MHz, other behaviour is seen in the low frequency studies performed by \cite{Kuzmin2007} and \cite{Geyer2017a}. \cite{Kuzmin2007} identified a $\tau$--DM relation of $\tau = 60(\rm{DM}/100)^{2.2}$ ms at 100~MHz, and \cite{Geyer2017a}, whose results for $\tau$ also corresponded well to this equation, measured a distribution of values of $\alpha$ that is systematically shifted to lower values in comparison to ours. An explanation for this may be that low frequency pulsar observations are probing a different scattering environment. One aspect of this is that lower frequency pulsar emission probes a wider region of space than that at higher frequencies since it is scattered more strongly, as shown in fig. 1 of \cite{Cordes2016}. Applying the same isotropic model may therefore result in different measurements of $\alpha$ at different frequencies, something not measurable for the frequency range of the data presented here but meriting further exploration. A further consideration is the effect of distance. Scattering analyses performed at low frequencies focus on nearby pulsars, due to the extreme scatter broadening that occurs in high DM pulsars. It is interesting to consider the effect of anisotropy in this context. First, having compared anisotropic model fits with the isotropic fits presented in this paper and found them to be generally unsuccessful, we see no evidence of anistropic scattering in the sample presented in this paper. Secondly, for nearby pulsars, it is likely that the pulsar emission only passes through a single or a very small number of screens, so that the scattering behaviour observed may bear hallmarks of anisotropy. There is evidence of anistropic scattering in low frequency surveys, e.g. \cite{Geyer2017a}, and using an isotropic model to identify the scattering parameters where such anisotropy is evident then results in under-estimation of $\alpha$, as described by \cite{Geyer2016b}. This may be another factor contributing to the smaller values of $\alpha$ at lower DMs seen in Fig. \ref{fig:alpha_DM_context}: pulsars with smaller DMs are likely to be nearby and hence an anisotropic model may be more appropriate. By contrast, the emission from our high DM pulsars is likely to have passed through multiple scattering screens, all of which may have different levels of anisotropy oriented along different axes. The net scattering effect observed will therefore approximate isotropy, as observed in our data.

\section{Conclusions}
\label{sec:conc}

As part of the TPA project we have identified 205 single-component profiles that appear to be scattered, out of a total of 1164 pulsars 
observed with the MeerKAT telescope. Of these we have obtained good scattering model parameters, $\tau$ and $\alpha$, for 84 pulsars, the largest sample of scattering parameters observed with a single telescope over a continuous bandwidth. Through this we have also obtained estimates of these pulsars' dispersion measures. The preferred choice of DM depends on the purpose for which dedispersion is being applied, and these DMs may be used to obtain accurate alignment of intrinsic pulse profiles for these pulsars, independent of scattering. 

Our investigations have identified and highlighted a key cause of inaccurate scatter modelling as covariance of model parameters for the intrinsic pulse profile and the scattering timescale. We have investigated the regimes under which such covariance is likely to cause problems and identified the effects of poor modelling as a tendency to over-estimate $\tau$ and under-estimate $\alpha$. Putting our results into the context of previous work, we propose that high-frequency, high-DM scattered pulsar observations are in general well modelled by the simple assumption of isotropy. Our measured average scattering spectral index, $\langle\alpha\rangle = 4.0 \pm 0.6$, agrees with both of the simplest theoretical models, 4 (thin screen) or 4.4 (Kolmogorov continuum), to within 1-$\sigma$ uncertainty. 
More complex cases, in particular pulsars where anisotropic scattering is measurable, are observable at lower frequencies, whereas the large distances to the pulsars observable at our frequency band will result in an average scattering effect that is net isotropic. 
Our results show no evidence of flatter power laws at lower frequencies that might indicate a truncated scattering screen or the effect of the inner scale.

This large body of measured scattering parameters provides a strong basis from which small variations can be determined accurately. For example, extreme scattering events \citep{Walker2007} could be detected and studied by investigating how $\alpha$ varies in observations of the same pulsars at different times. The software we developed to perform the scattering analysis is called \mbox{\sc SCAMP-I} \mbox{(SCAtter Modelling for Pulsars - I)}. It is suitable for time domain modelling of single-component scattered pulse profiles and is available for use at \mbox{https://github.com/pulsarise/SCAMP-I }

\section*{Acknowledgements}
The MeerKAT telescope is operated by the South African Radio Astronomy Observatory, which is a facility of the National Research Foundation, an agency of the Department of Science and Innovation. MeerTime data is housed and processed on the OzSTAR supercomputer at Swinburne University of Technology with the support of ADACS and the gravitational wave data centre via AAL. LO acknowledges funding from the UK Science and Technology Facilities Council (STFC) Grant Code ST/R505006/1. AK also acknowledges funding from the STFC consolidated grant to Oxford Astrophysics, code ST/000488. RMS acknowledges support through Australian Research Council Future Fellowship FT190100155.

\section*{Data Availability}
The data underlying this article will be shared on reasonable request to the corresponding author.




\bibliographystyle{mnras}
\bibliography{Scattering} 



\appendix

\section{Theoretical sources of poor $\tau$ modelling}
\label{app:theory}

We looked to quantify theoretically the circumstances under which we would expect $\tau$--$\sigma$ covariance or spectral flattening (as described in sections \ref{sec:covar} and \ref{sec:nonPL}) to appear in our scatter modelling. 
We consider the functional form of the model we apply---the convolution of a Gaussian with an exponential decay---which is as follows: 
\begin{equation}
    I(t) = \frac{1}{2}\exp\left({-\frac{t}{\tau} + \frac{\sigma^{2}}{2\tau^{2}}}\right)\left(1 + \textrm{erf}\left(\frac{t}{\sqrt{2}\sigma} - \frac{\sigma}{\sqrt{2}\tau}\right)\right).
    \label{eq:modelfunc}
\end{equation}
As $\sigma$ tends to 0, this equation tends to pure exponential decay, as expected. The extent to which $\sigma$ is relevant to the shape of the profile depends on the error function term $x = \left(\frac{t}{\sqrt{2}\sigma} - \frac{\sigma}{\sqrt{2}\tau}\right)$. For a given ratio of $\sigma/\tau$, the Gaussian part of the model will play a large contribution in the overall model shape up to some time $t_{0}$, after which the exponential decay will dominate the pulse profile shape. We define $t_{0}$ as the time that, for a given ratio of $\sigma$ and $\tau$, gives $x = 2$. This is based on the shape of the error function, since at $x = 2$ the functional form starts deviating strongly from $\textrm{erf}(x) = 1$. 

\begin{figure}
    \includegraphics[width=\columnwidth]{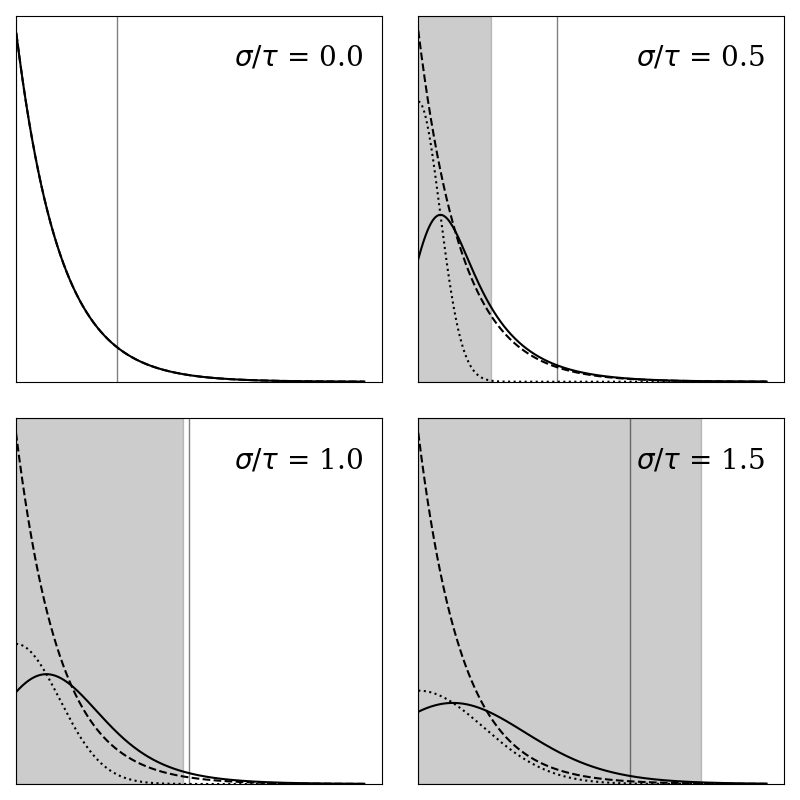}
    \caption{Plots indicating how the Gaussian and exponential components of our scattering model contribute to the overall model shape, and how that changes with increasing $\sigma$/$\tau$. Dashed line: the exponential decay function described by equation \ref{eq:iso}. Dotted line: a Gaussian function. Solid line: the convolution of these two, described by equation \ref{eq:modelfunc}. Grey vertical line: this marks the phase point at which the combined model intensity (solid line) drops to 10\% of its maximum. Grey shading: this indicates the region where the Gaussian (dotted line) is contributing strongly to the overall profile shape (solid line). The definition of a strong contribution is given in the text. Each subplot shows the same set of functions: the changed curve shapes results from the changed ratio of $\sigma/\tau$, which is marked in the top right corner of each subplot.}
    \label{fig:sigtau_theory}
\end{figure}

In Fig. \ref{fig:sigtau_theory} we plot equation \ref{eq:modelfunc}, along with its constituent Gaussian and exponential components, for 4 different values of $\sigma/\tau$. For each, we shade that part of the profile where $t < t_{0}$. As we can see in the figure, a small $\sigma/\tau$ results in small $t_{0}$, so that the majority of the pulse profile shape ($t > t_{0}$) is dominated by exponential decay and $\tau$ can be determined easily. 
If $\sigma/\tau$ is large, 
then we may expect to see covariance between the measured values of $\sigma$ and $\tau$. 
Replacing our Gaussian model with a profile observed at a high frequency, that is considered to be unaffected by scattering, has previously been common practice, however not accounting for profile evolution brings its own inaccuracies. 

We define the Gaussian contribution as too large when $t_{0}$ coincides with the point at which the total intensity has dropped to 10\% of its peak. We mark the 10\% intensity point with a vertical line in Fig \ref{fig:sigtau_theory}. This means that the cut-off ratio of $\sigma/\tau$ for which the underlying pulse profile shape interferes too much in the scattering measurements to be able to reliably separate $\sigma$ from $\tau$, is $\sigma/\tau \sim 1$. 

\begin{figure}
    \includegraphics[width=\columnwidth]{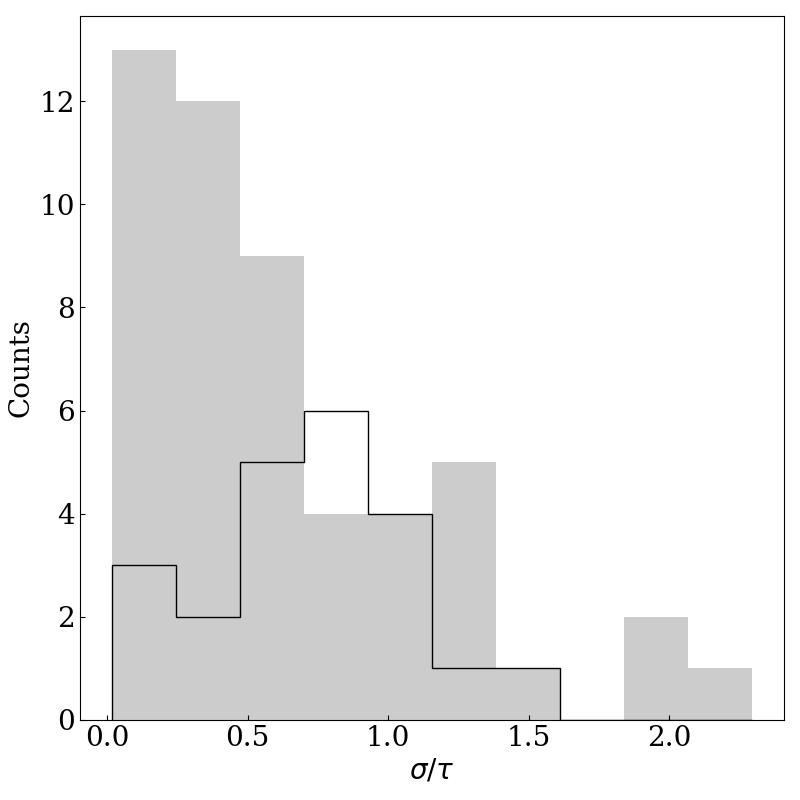}
    \caption{Histograms of the ratios of measurement results $\sigma$/$\tau$. Grey: values for the highest frequency profile fits for the pulsars with a standard power-law $\tau$-frequency relationship. Transparent, black-edge: values for the highest frequency profile fit that is still following the power-law relationship for those pulsars exhibiting tau-flattening.}
    \label{fig:sigtau_hist}
\end{figure}

In Fig. \ref{fig:sigtau_hist} we show the histograms of values of $\sigma/\tau$ for two groups of pulsars. For the pulsars for which we are confident of the fit at all frequencies, we have calculated $\sigma/\tau$ at the highest observing frequency, and plotted the values as a shaded grey histogram. For the pulsars showing the spectral flattening behaviour, we have taken $\sigma/\tau$ at the highest frequency for which the values are still following power law behaviour: we treat this as the cut-off point. The histogram of these values is plotted in Fig. \ref{fig:sigtau_hist} as a stepped histogram. We see that the majority of pulsars with no spectral flattening have $\sigma/\tau < 0.7$, in accordance with our theoretical estimates that the parameters can be measured accurately for smaller values of $\sigma/\tau$. By contrast, the histogram of cut-off values peaks at $\sigma/\tau \sim 0.8$, implying that the flattening we see at higher frequencies is indeed due to the loss of ability to separate the intrinsic profile shape from the exponential scattering. Our theoretical estimate of $\sigma/\tau = 1$ being the cut-off for successful modelling is intended only to be indicative, since it takes no account of the variety of signal-to-noise ratios or intrinsic profile shapes of our observations. 
It is therefore unsurprising that we see large spreads in both histograms in Fig. \ref{fig:sigtau_hist}. 
In particular, the left-hand side of the stepped histogram is consistent with explaining those pulsars which, like PSR J1653--4249, may have extra components that are altering the scattering behaviour of the pulsar in comparison to what we might expect.

These results indicate the extent to which individual measurements of $\tau$ of single profiles are vulnerable to many sources of bias and error. Scattering properties of pulsars can only be characterized reliably in cases where scattering measurements can be performed across large frequency bands. Further, the scattering results for a single pulsar, and the science that can be inferred from them, are best understood in the wider context of all of the other pulsars observed and analysed with the same method.

\section{Simulation: how extra profile components affect scattering timescale measurement}
\label{app:sim}

\begin{figure}
    \includegraphics[width=\columnwidth]{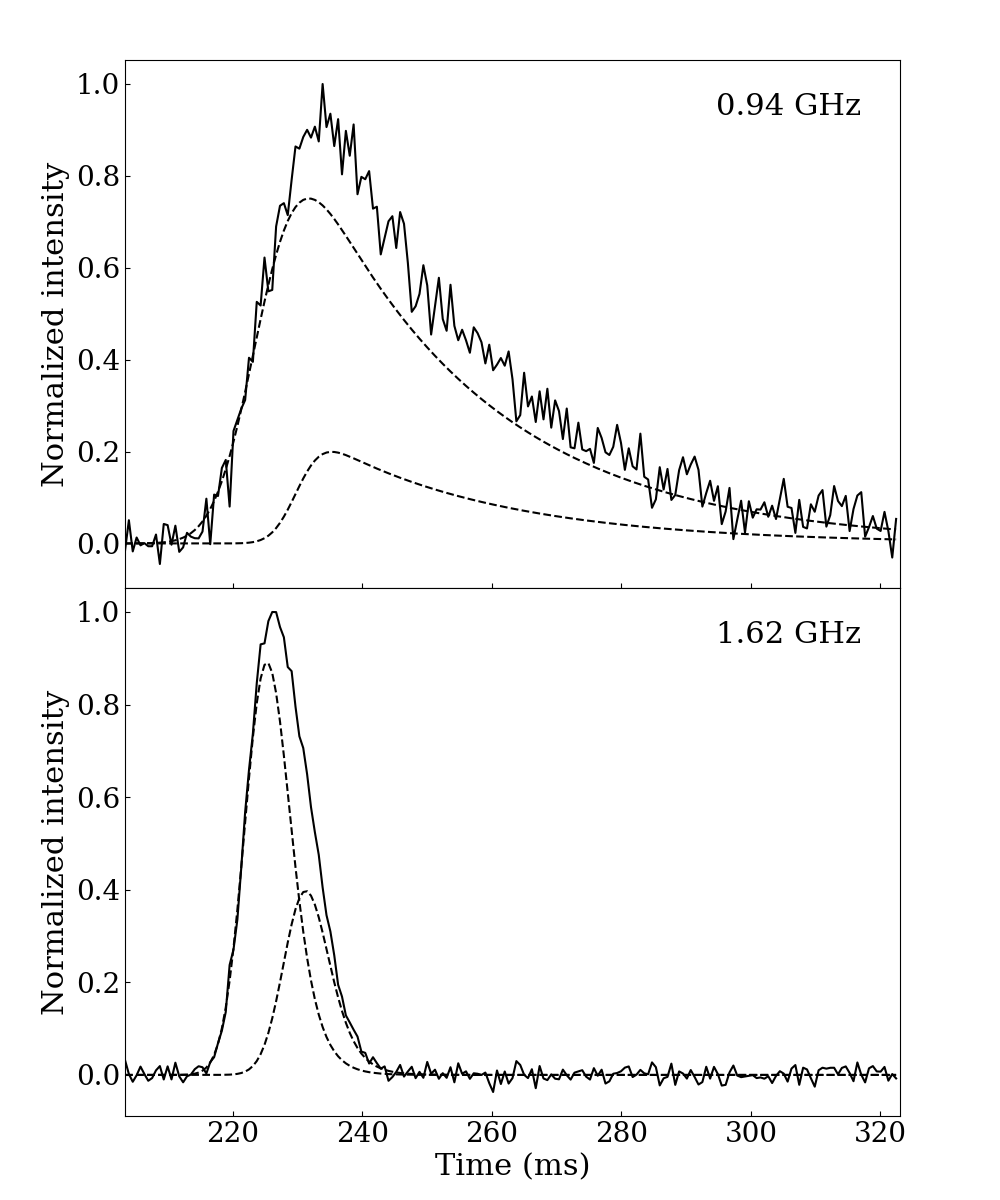}
    \caption{Profiles for a simulation of a scattered double-component pulsar, showing the lowest (top) and highest (bottom) frequency simulated profiles respectively. Solid line: the simulated profile. Dashed lines: the two noiseless components that comprise the simulation. These are added together and then noise is added to generate the simulated profile. The subplots are zoomed in to show only the on-pulse region. }
    \label{fig:sim_prof}
\end{figure}

We attempted to replicate the output of the scatter modelling process of PSR~J1653--4249 with simulated data, to test whether the presence of a hidden second component in the pulse profile could be responsible for the spectral flattening behaviour seen in this pulsar's modelling results.
We simulated a pulse profile made up of two Gaussian components: a large main component and a smaller, narrow secondary component that sits to the right of the main peak. 
On top of this, we introduced profile evolution: the width of the main component decreases with increasing frequency according to a power law plus constant relationship \citep{Thorsett1991}, and the flux spectral index of the secondary component is flatter than that of the main one, reflecting what is commonly seen in observations. We defined the scattering timescale at the lowest frequency to be the same as that measured for the data. We then applied a spectral index of $\alpha = 4.4$ to $\tau$ to obtain its value at other frequencies. We defined the height, width and position the second component such that the overall profile shapes of the simulation appear similar to those of the real pulsar. Fig. \ref{fig:sim_prof} shows how the two components combine to make the simulation profile shapes at the lowest and highest frequencies. 

\begin{figure}
    \includegraphics[width=\columnwidth]{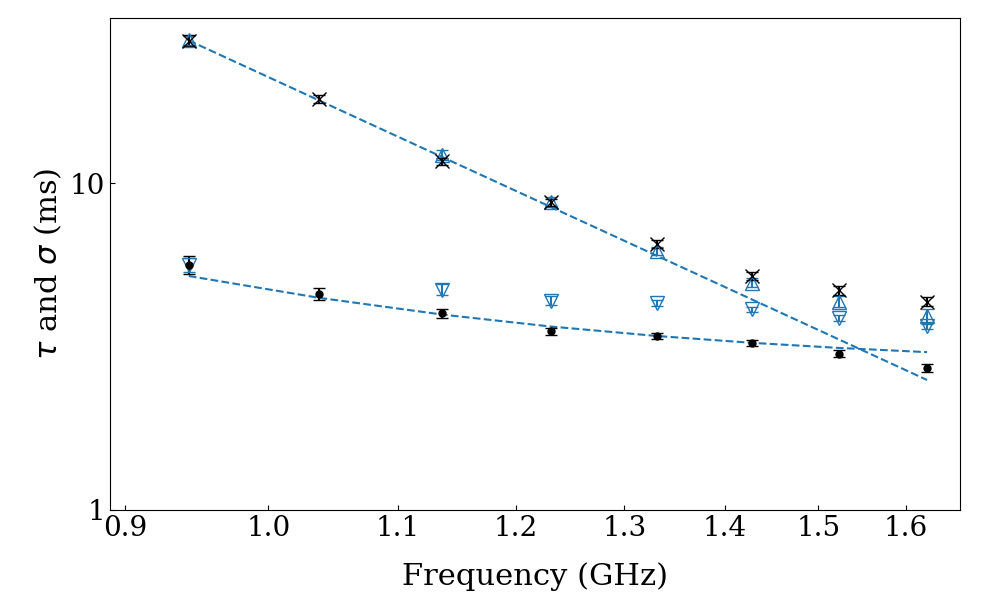}
    \caption{Log-log plot of scattering timescale $\tau$ and intrinsic Gaussian standard deviation against frequency, similar to Fig. \ref{fig:combi_tausigfreq}. Black: modelling results for PSR~J1653--4249, showing $\tau$ (crosses) and $\sigma$ (points). Blue triangles: modelling results for a simulated scattered double-component pulsar, showing $\tau$ (upright) and $\sigma$ (upside-down). Blue dashed lines: input values of $\tau$ and $\sigma$ used to generate the simulated pulse profiles.}
    \label{fig:sim_extracompt}
\end{figure}

Performing the MCMC fit on this simulation, we obtained results for $\tau$ and $\sigma$ that we compare to those of the data in Fig. \ref{fig:sim_extracompt}. The effect of the second component is as expected: its presence does little to alter the fit parameters at the low frequency end, where they are recovered well, but at higher frequencies we see a flattening off in the spectral behaviour of both $\tau$ and $\sigma$ that mimics what we witnessed in our measurements for PSR J1653--4249. This lends credence to our choice, for pulsars like PSR J1653--4249, to keep only those values of $\tau$ where a power law is still being followed, and infer an $\alpha$ from those.

\bsp	
\label{lastpage}

\end{document}